# Pour une interopérabilité sémantique en éducation : les modèles normatifs de l'ISO/IEC JTC1 SC36

Mokhtar Ben Henda
MICA/GRESIC, Université de Bordeaux 3
Mokhtar.benhenda@u-bordeaux3.fr

**Résumé :** La sémantique des contenus est l'une des constituantes essentielles des modèles de dispositifs pédagogiques innovants. Elle se construit progressivement sur la base d'efforts normatifs accomplis par différents acteurs dans les domaines de l'industrie technologique, des télécommunications, de l'informatique, de l'ingénierie linguistique, des sciences de l'information et de la documentation etc. La sémantique dans les réseaux et les systèmes d'information numériques représente, en fait, un maillon avancée d'un processus long de traitement de l'information numérique dans lequel le travail terminologique occupe une part importante. Ce processus est également très consolidé par une large panoplie de normes et de standards qui assurent des niveaux très élevés d'interopérabilité technique, organisationnelle et sémantique.

**Mots clés :** *Normalisation, e-Learning, terminologie, sémantique, interopérabilité*

## 1. Cadre général

Avec l'avènement du Web 3.0, les systèmes d'information numériques progressent vers un nouveau palier de performance : les réseaux sémantiques. Bien qu'avec un certain retard, le monde de l'éducation est en train de s'approprier les acquis techniques, systémiques et procédurales de ces nouveaux systèmes et de converger vers des nouvelles formes de gestion et de capitalisation des connaissances. La sémantique des contenus est l'une des constituantes essentielles sur lesquelles sont conçus les nouveaux modèles de dispositifs pédagogiques. Elle se construit progressivement sur la base des efforts normatifs réalisés par différents acteurs dans les domaines de l'industrie technologique, des télécommunications, de l'informatique, de l'ingénierie linguistique, des sciences de l'information et de la documentation etc. La sémantique des réseaux représente en fait, un maillon avancée d'un processus long de traitement de l'information numérique depuis le codage informatique jusqu'au balisage sémantique et la fouille des textes en passant par la conception des schémas de métadonnées pour la description des ressources et la définition des vocabulaires pour créer des ontologies et des cartes conceptuelles. Cette chaine est aujourd'hui très consolidée par une large panoplie de normes et de standards qui assurent des niveaux très élevés d'interopérabilité, technique, organisationnelle et sémantique.

L'e-Learning entame ainsi un long processus de normalisation pour atteindre des niveaux optimums d'adaptation, d'intégration et de convergence des ses moyens et ressources éducatives. Son objectif est d'asseoir un cadre opérationnel d'interopérabilité générale (acteurs, outils, services et contenus) et de fonder des mécanismes de partage de ressources à échelle internationale. Ces efforts de normalisation touchent un grand nombre de facettes relatives à l'apprentissage par les technologies éducatives. La sémantique est l'une de ces facettes autour de laquelle travaillent plusieurs organismes et structures internationales de normalisation. Des acquis normatifs pour l'interopérabilité des systèmes sont déjà entrés en vigueur par des structures régionales ou internationales comme le W3C, l'ISO ou le CEN. L'éducation est en phase de développer ses propres références qui sauraient établir une forme de gouvernance mondiale de l'enseignement par les TIC.

Ce papier présente les références normatives les plus importantes en cours d'usage ou de développement, conçues et adoptées par des structures de normalisation comme l'ISO/TC37[1] ou l'ISO/IEC JT1 SC36[2]. Un état de l'art est exposé avec des réflexions sur les contraintes et les perspectives d'une progression vers des réseaux sémantiques interopérables en éducation grâce aux standards technologiques et aux normes d'interopérabilité.

## 2. Les schémas de métadonnées, un prérequis de l'activité sémantique

À l'instar de tout système d'information requérant un mécanisme de description formelle de ses ressources par des éléments de métadonnées,

---

[1] ISO/TC37 : Comité technique n°37 de l'ISO, chargé de la normalisation du vocabulaire dans tous les domaines.

[2] ISO/IEC JTC1 SC36 : Sous comité n°36 du comité joint de l'ISO et de la Commission électrotechnique internationale, chargé de la normalisation des technologies éducatives et de l'e-Learning

l'univers de l'enseignement se démarque aussi par la conception et la mise en place de ses propres modèles de description des ressources et dispositifs d'enseignement et d'apprentissage. Les schémas de métadonnées pédagogiques sont à l'origine d'un long processus de normalisation du champ éducatif et plus particulièrement universitaire.

Les métadonnées sont devenues l'un des mécanismes essentiels pour référencer, localiser et restituer des ressources sur les réseaux. À l'instar du fonctionnement des modèles d'indexation des ressources électroniques par le protocole OAI-PMH (*Open Archives Initiative/Protocol for Metadata Harvesting*) dans les archives ouvertes ou le schéma de métadonnées *Dublin Core* sur Internet, on commence à voir apparaître des réservoirs d'objets d'apprentissage constitués de matériaux pédagogiques multidisciplinaires. Grâce à des métamodèles de description par métadonnées pédagogiques comme le modèle LOM (*Learning Object Metadata*) et des standards de construction d'agrégations de contenus comme SCORM (*Sharable Content Object Reference Model*), ces réservoirs deviennent de plus en plus partageables et interopérables. Ils deviennent de plus en plus sujets à des traitements d'ordre sémantique pour faciliter leurs repérages, restitutions et réutilisation. Le principe de référencement normalisé s'est même généralisé à d'autres aspects du fonctionnement des structures académiques. Il concerne, entre autres, les offres de formations dans les structures universitaires dans un objectif de les rendre identifiables, localisables et interopérables.

Les schémas de métadonnées sont, de ce fait, un cadre structurant de la description des entités par des concepts et des mots clés significatifs déterminés par l'auteur des ressources ou par les acteurs exécutant cette fonction de référencement (*e.g.* bibliothécaires, documentalistes). Cette fonction obéit à une réglementation et à une normalisation parfois stricte dans le choix des termes de description. C'est l'univers de la terminologie et du vocabulaire contrôlés qui associe toutes les ressources traitant des mêmes thèmes sous les mêmes identifiants conceptuels et terminologiques. Des outils bibliographiques et documentaires existent depuis longtemps dans un grand nombre de disciplines. Thésaurus, listes d'autorité, liste de vedette matières etc. sont parmi les outils classiques des documentalistes indexeurs des ressources documentaires dans les systèmes classiques des bases et banques de données documentaires ou des catalogues des bibliothèques accessibles en ligne. Avec le réseau Internet, la couche sémantique est beaucoup plus complexe. On évoque désormais des nouvelles catégories d'outils normalisés comme les ontologies et les *Topic Maps* comme moyens d'organisation et de gestion des ressources numériques sur les réseaux. Vu la complexité de leur élaboration sur un plan technique ou sémantique, ces outils ont tendance à se spécialiser dans des disciplines ou des branches de disciplines. L'*e-Learning* est l'une des branches qui cherche à se doter de ses propres outils pour évoluer graduellement depuis la terminologie et le vocabulaire spécialisé jusqu'à la sémantique des contenus.

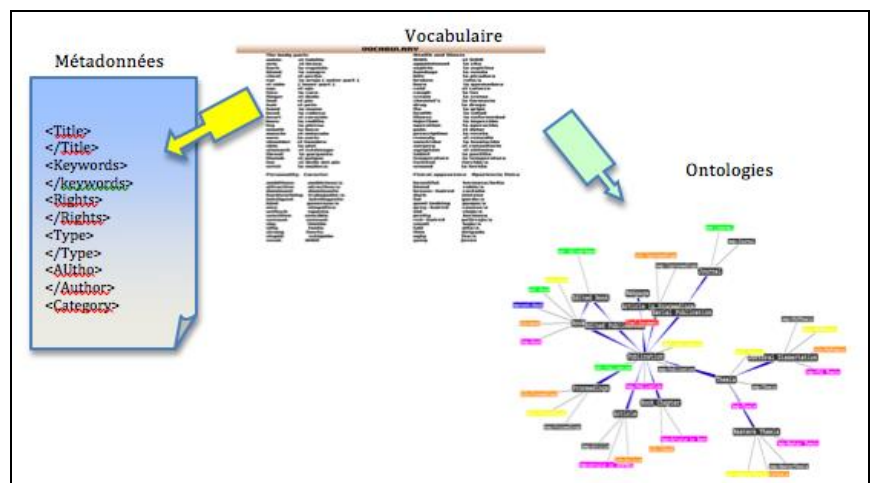

*Fig. 1 : articulations entre métadonnées, vocabulaires & ontologies*

## 3. Le champ terminologique

La question du vocabulaire est stratégique à plus d'un titre. La terminologie normalisée est l'un des fondements des réseaux sémantiques. Dans le contexte *e-Learning*, développer une terminologie normalisée du domaine est indispensable. Elle constitue l'alternative obligatoire pour les développeurs des schémas des métadonnées afin de résoudre les problèmes inhérents à la diversité des sources pouvant servir à renseigner certains éléments de métadonnée. À part les listes prédéfinies des valeurs fixes pour des éléments de métadonnées, comme les types documentaires ou les états de finition des ressources, certains éléments sont ouverts et donc associés à des registres terminologiques multiples. Le vocabulaire est également important pour les développeurs de contenus. Ceux-ci sont de plus en plus impliqués dans le référencement de leurs productions scientifiques ou pédagogiques par une terminologie qui doit répondre de plus en plus à des valeurs conceptuelles et sémantiques communes. Pour faire face à l'ouverture exponentielle des réseaux et des systèmes d'information, et aux risques de la dispersion des ressources, la normalisation terminologique agit encore une fois en tant qu'agent fédérateur d'initiatives et de modèles. Le monde de l'*e-Learning* œuvre depuis quelques temps au développement de ses propres outils et modèles terminologiques pour la construction d'ontologies spécialisées *e-Learning*.

Dans la suite de ce document, nous aborderons les grandes lignes d'une approche normative du domaine terminologique en rapport avec la sémantique de l'*e-Learning*. Nous décrirons ensuite les méthodes et les techniques que nous avons entreprises pour procéder à la production de vocabulaire *e-Learning* multilingue et normalisé selon les normes en vigueur spécifiques au TC37et au SC36.

### *La terminologie au milieu d'un champ notionnel de concepts*

Avant d'entrer dans les détails des champs sémantiques et normatifs, il serait utile de clarifier quelques concepts dont la dénomination et le sens prêtent souvent à confusion. À moins d'être expert, on a souvent tendance à confondre le sens du vocable « terminologie » avec ceux de « vocabulaire », de « lexique » et même d'« ontologie ». C'est aussi pareil dans l'usage des expressions « Mot », « Terme », « Concept » et « Notion » alors que la différence entre eux est fondamentale.

Selon la définition du dictionnaire de l'Académie française, 8ème édition, un vocabulaire est une « *liste de mots, rangés habituellement dans l'ordre alphabétique et accompagnés d'une explication succincte. […] Il se dit aussi de l'Ensemble des mots employés par un peuple, par un groupe, par un écrivain, etc. […] Il se dit encore des Mots qui appartiennent particulièrement à une science, à un art* ». Un vocabulaire est donc une liste de « vocables » (l'expression n'est pas courante) ou de « Mots » qui ont été déterminés pour des besoins immédiats d'information ou de communication. Un vocabulaire peut toutefois être défini et contrôlé par une communauté pour des besoins pratiques, sans pour autant, que le sens ou la relation entre les mots soient définis. Dès qu'il est organisé et structuré en hiérarchie, un vocabulaire se spécialise et devient une taxonomie avec des liens précis entre les « Mots » (Polguère, 2003).

Le « Concept » ou la « Notion » est à un niveau d'abstraction plus élevé que le « Terme » ou le « Mot ». Il s'agit, de façon générale, d'une idée abstraite ou d'un concept mental qui se distingue aussi bien de la chose représentée que du mot ou de l'énoncé verbal qui la représente.

L'ontologie, pour sa part, n'est pas facile à définir non plus, car elle prend racine dans des contextes très différents comme la philosophie, la psychologie, la linguistique, l'informatique ou l'intelligence artificielle. On pourrait toutefois partir de la définition très courante de Gruber qui identifie l'ontologie comme une spécification explicite et formelle d'une conceptualisation partagée (Gruber, 1993). À la différence d'un vocabulaire, donc des mots ou vocables, l'ontologie cherche à représenter le sens des concepts et les relations qui les lient dans un réseau sémantique.

Depuis l'apparition du concept des réseaux sémantiques, les ontologies ont connu un regain d'intérêt traduit par une prolifération d'ontologies de domaines. La sémantique des réseaux est largement définie dans les relations entre les termes constituant une ontologie. Il faudrait rappeler ici que la sémantique se distingue aussi de la terminologie par la fait qu'elle se préoccupe de la relation entre la dénomination et le signifié, alors que la terminologie s'intéresse d'abord à la relation entre l'objet réel et la notion qui le représente. Les relations entre deux termes représentant des concepts sont généralement associées à l'interprétation humaine qui en est faite. Mais, dans les réseaux sémantiques électroniques actuels, une sémantique formelle peut permettre des calculs automatiques pour vérifier si une cohérence existe entre des informations inscrites pour décrire une connaissance. D'un point de vue représentation, les connaissances et leurs relations sont désormais classées par taxonomies, classifications ou thesaurus. Ils prennent forme de graphes conceptuels ou de réseaux sémantiques.

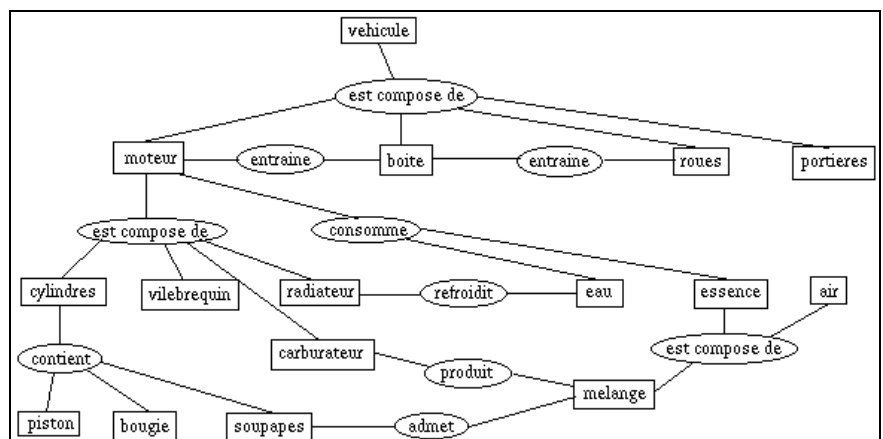
*Fig. 2 : Exemple d'un graphe conceptuel (ontologie)*

**Terminologie et normalisation : le référentiel ISO du TC37**

Les institutions de normalisation sont de plus en plus conscientes de l'urgence de systématiser l'appréhension de concepts définis en consensus. Elles recommandent instamment aux modérateurs et responsables de chaque instance de normalisation de désigner un groupe de travail spécifique chargé d'élaborer une terminologie du domaine. Il est cependant opportun de remarquer que les experts en normalisation dans tous les métiers ne disposent pas des mêmes niveaux de sensibilités par rapport au besoin de normaliser les terminologies de leur domaine de spécialité. Malgré l'existence de normes internationales qui recommandent une pareille initiative (ISO TC37), les terminologies produites par les structures de l'ISO sont techniquement et méthodologiquement très variées. Leurs handicaps majeurs sont sans doute l'irrégularité de leur actualisation et mise à jour, puis surtout le manque de coordination interdisciplinaire entre des terminologies spécialisées. La mise en place de méthodes de développement terminologique, rigoureusement normalisées est une voie d'avenir pour assurer une interopérabilité terminotique conforme aux défis de la mondialisation et aux avancées vers un véritable Web sémantique. C'est justement le souci d'assurer cette interopérabilité terminologique qui définit l'un des axes de travail du Sous Comité 36 afin de produire une terminologie *e-Learning* normalisée et multilingue.

L'une des normes en cours de développement par le TC37/SC1 est la norme « *ISO 860:2007 -- Travaux terminologiques -- Harmonisation des concepts et des termes* ». Elle définit une approche méthodologique qui permet l'harmonisation des concepts, des systèmes de concepts, des définitions et des termes. Elle s'applique à l'élaboration de terminologies harmonisées, au niveau national ou international, dans un contexte monolingue ou multilingue. Le suivi de cette norme, dont la révision est encore au stade de NP (*New Work Project*), est fondamental pour l'harmonisation des vocabulaires produits et adoptés par différentes structures produisant et utilisant des normes de vocabulaire.

Trois autres normes clés du TC37 sont aussi stratégiques à cet effet :

La norme « *ISO 704:2009 -- principes et méthodes* » est à la base de tout travail terminologique. Elle établit et harmonise les principes fondamentaux et les méthodes permettant d'élaborer et de compiler des terminologies, qu'il s'agisse d'activités menées dans le cadre de la normalisation ou non. Elle décrit aussi les liens établis entre les objets, les concepts et leur représentation par des terminologies.

La norme « *ISO/IEC 16642:2003 -- (TMF : Terminological Markup Framework)* » précise un cadre destiné à fournir des orientations sur les principes de base pour représenter les données terminologiques.

La norme « *ISO/IEC 10241:1992 -- Normes terminologiques internationales -- Élaboration et présentation* ». Cette norme internationale établit des règles à suivre dans l'élaboration et la présentation de terminologies normalisées. Elle propose les directives de présentation normalisée des listes de vocabulaires.

L'intérêt des normes du TC37 est qu'elles ouvrent le cadre d'interopérabilité entre différentes bases de données terminologiques. Cette interopérabilité est d'abord assurée par l'identité de méthode et du mode de description. Plusieurs bases de données terminologiques qui s'appuient sur la norme « *ISO 704:2009 -- principes et méthodes* » et la norme « *ISO/IEC 12620:1999-Aides informatiques en terminologie -- Catégories de données* », seront déjà assurées d'avoir constitué des ressources terminologiques qui peuvent (au prix d'efforts informatiques importants mais possibles) être récupérées pour être interopérables entre elles. Dès lors, il devient clair que ces 10 dernières années, les efforts terminotiques ont été très importants et se sont focalisés sur le modèle TMF et l'adaptation des nouvelles versions de la norme « *ISO/IEC 12620:1999* » à ces nouvelles méthodes.

## 4. Construire des terminologies normalisées : les prémices d'un cadre sémantique e-Learning

Dans le monde de l'éducation, le développement de référentiels terminologiques concernant les technologies, les descriptions d'institutions, les disciplines, les modes de certification (diplômes, niveaux), les styles pédagogiques, les contextes juridiques etc. sont de plus en plus nécessaires pour permettre la circulation internationale des ressources d'enseignement et de formation. La généralisation constante des nouveaux modes d'enseignement par les TICE fait de l'enseignement, de la formation et de l'apprentissage des valeurs universelles de développement. La mise en place de ressources terminologiques libres d'accès, qui soient un véritable « bien public », est, dès lors, indispensable. Or, l'offre actuelle de terminologies normalisées est très inappropriée car elle est constituée de ressources souvent produites selon des méthodes orientées sur les termes et non les concepts. La question des méthodes d'élaboration des corpus terminologiques allait, de ce fait, être déterminante. Le débat s'est concentré ces dernières années sur les deux approches les plus connues dans la conception de terminologies normalisées. Il s'agit des deux approches méthodologiques onomasiologique et sémasiologique qui relèvent généralement du domaine de la lexicologie.

L'approche sémasiologique (du grec *sêma* 'signe, sens') étudie en général la structure des lexiques. Puisque les dictionnaires partent généralement de la forme du mot et énumèrent ses différentes significations, la sémasiologie décrit la polysémie d'un mot et les types de relations existant entre ses différentes significations tout en analysant leur rapport avec les entités concrètes du monde et avec celles du domaine conceptuel.

La démarche de l'approche onomasiologique (du grec *ónoma* 'nom'), consiste à élaborer un système théorique qui pourrait rendre compte, dans un cadre d'abstraction totale, de toutes les valeurs qu'une notion pourrait avoir. Cela donne lieu à une construction de tableaux à plusieurs entrées qui sont ensuite confrontés aux différents systèmes morphologiques pour les faire coïncider à des langues naturelles. C'est donc le concept qui constitue le point de départ.

Cette fonction est généralement assurée par les thésaurus qui suivent une approche typiquement onomasiologique puisqu'elle va du concept ou de la signification aux différents synonymes du même concept. Il est donc du ressort de l'onomasiologie de traiter des mots qui ont une signification similaire, c'est à dire des relations de synonymie.

Le domaine terminologique s'oriente traditionnellement vers une démarche onomasiologique en s'intéressant prioritairement à des notions (concepts) et aux mots ou expressions qui les nomment. Elle les traduit ensuite en termes d'une classification conceptuelle. La tendance terminotique est désormais dans l'étude des modes d'introduction des termes nouveaux dans les textes et les discours et non seulement dans celle des termes isolés. Pour ce qui est des documentalistes, des spécialistes du *knowledge management*, des ingénieurs ou des chercheurs analysant des processus techniques ou un phénomène scientifique, on peut noter qu'ils ont généralement tendance à partir des concepts pour pointer sur un $2^e$ niveau de termes dans des langues naturelles, voire des langues de spécialité. Aujourd'hui, la communauté des terminologues, mais aussi celle des lexicographes, s'est entendue pour normaliser une démarche unique, celle qui part des concepts pour déterminer les termes.

C'est de cette logique que découle les arguments de la communauté des normalisateurs terminologues et lexicographes pour prendre comme seuls valides, les principes et les méthodes correspondant à la démarche onomasiologique de la norme « *ISO 704 - Terminologie : principes et méthodes* ». Ils ont ensuite normalisé un catalogue ouvert de « catégories de données » apte à définir des données terminologiques ou lexicographiques. C'est la norme « *ISO/IEC 12620:1999-Aides informatiques en terminologie -- Catégories de données* ». Enfin, ils ont normalisé un cadre commun de mise en œuvre terminotique à même d'assurer l'interopérabilité et la réutilisabilité des ressources terminologiques indépendamment des diverses banques de donnes terminologiques. Ce cadre commun, le TMF (*ISO/IEC 16642:2003-Applications informatiques en terminologie -- Plate-forme pour le balisage de terminologies informatisées*), nécessite bien sûr que ces différentes bases respectent le métamodèle XML de TMF, ou exige que les ressources terminologiques soient reformatées selon ce même modèle.

Bref, la terminotique n'a réellement bouclé ses méthodes que grâce au progrès de l'information structurée avec SGML et surtout avec XML. Les deux ont permis d'organiser toutes les banques de données terminotiques (notamment les terminologies multilingues), aussi bien en conciliant la logique lexicographique avec la logique terminologique qu'en permettant surtout de concevoir un schéma terminotique unique, en l'occurrence *TMF (Terminological Markup Framework)*. En effet, TMF permet d'assurer l'interopérabilité de toutes les banques terminologiques ou lexicographiques selon les multiplications des nombres de langues que l'on peut souhaiter. L'interopérabilité des terminologies dans le futur Web sémantique exigeait notamment un tel choix normatif. Les mécanismes du TMF, et la normalisation des catégories de données permettent précisément de servir de base à la réalisation modulaire d'ontologies, elles même en cours de standardisation par le W3C et le langage OWL (*Web Ontology Language*).

C'est sur la base de toutes ces considérations, techniques et procédurales, que le SC36 œuvre depuis une décennie pour tenter de fournir, non pas un œuvre terminologique exhaustive de l'*e-Learning,* mais plutôt pour mettre en application les procédures normatives de conception et développement de ces corpus et de leur appropriation par les experts internationaux, chacun dans son contexte, sa langue et sa culture. Comment se définit alors le cadre général dans lequel agit le SC36 ? Dans quelle perspective répond-il à une demande concrète des pays membres et du monde entier ? Quels sont ses outils, ses choix et ses prérogatives pour y parvenir ?

## 5. Le cadre de développement terminologique de l'ISO/IEC JTC1 SC36

Nous avons décrit en détail le contexte de travail du SC36 dans plusieurs documents antérieurs (Hudrisier & Ben Henda, 2008 ; Ben Henda, 2008). Mais pour des raisons d'actualité et de cohésion avec la finalité de ce document, celle de focaliser sur la dimension sémantique, nous préférons redessiner ici les contours de ce cadre normatif.

Depuis sa création en 1999, le SC36 s'est conformé aux règles de la majorité des instances normatives en créant un groupe de travail chargé de la terminologie du domaine que l'auteur de ce document coordonne depuis 2009. Il s'agit du groupe de travail n°1 sur la terminologie e-Learning (WG1). L'une des tâches à laquelle il devait faire face est la définition d'une approche de travail qui conduit aux résultats escomptés.

### *Une convergence vers le travail sémantique*

Les activités du WG1 sont en réalité canalisées sur deux axes parallèles. D'une part, la production des listes terminologiques normalisées des technologies éducatives et de l'e-Learning et d'autre part le travail sur une représentation sémantique du champ de l'e-Learning par les graphes conceptuels.

### *La production terminologique*

Le SC36 a fait plusieurs choix et expérimenté plusieurs solutions pour le développement d'une terminologie conforme aux normes en vigueur. L'un des outils qui a permis un début de production est « Genetrix », une application francophone développée par André Le Meur, anciennement informaticien au Laboratoire CRAIE de l'Université de Rennes 2. Bien que propriétaire, cette application a été conçue autour des normes du TC37 et fonctionne selon les spécifications de la norme *Geneter*. Il s'agit d'un cadre de travail bien adapté pour l'édition, la transformation, l'évaluation et l'échange de données terminologiques sur la base des principes de la norme ISO 704. *Geneter* est un modèle générique qui signifie qu'il contient tous les éléments de données terminologiques de la norme ISO 12620 pour évaluer la qualité formelle, les variations interculturelles, etc. Le WG1 a pu produire avec *Genetrix* des corpus terminologiques normalisés dans plusieurs langues. Mais, suite à des contraintes d'adaptabilité technique de l'outil aux besoins d'une communauté d'experts internationaux du SC36, *Genetrix* a du être abandonné dans l'optique de développer un autre outil terminologique *Open source* en langage XML.

Aujourd'hui, le WG1 utilise des outils adaptés du monde de la bureautique générale pour compiler, produire et éditer des listes terminologiques normalisées. Des outils spécifiques sont en cours de conception dans des cadres de partenariat avec des structures engagées sur la voie de la terminologie et de la sémantique e-Learning comme l'Agence universitaire de la francophonie (AUF). D'autres langues (coréen, russe, chinois, japonais) se sont jointes pour proposer des équivalents linguistiques à la production existante en anglais et en français. Elles feront partie de la norme publiée « *ISO/IEC 2382-36:2008 -- Technologies de l'information — Vocabulaire — Partie 36 : Apprentissage, éducation et formation* ».

### La convergence sémantique

La deuxième voie par laquelle le WG1 prévoit étendre ses activités vers des aspects sémantiques, est celle des ontologies et de la représentation sémantique par des graphes conceptuels. Un outil a été testé et retenu pour servir d'instrument de développement des cartes conceptuelles. Il s'agit du logiciel libre *Cmap tools*. Cet outils a été retenu pour produire des cartes graphiques conceptuelles à partir des termes et définitions retenues dans la norme « *ISO/IEC 2382-36:2008* ». Le principe consiste à extraire les concepts de la norme en question et les associer par des relations sémantiques. Ces relations sont normalement définies selon les directives de la norme « *ISO 10241:2003 -- Normes terminologiques internationales -- Élaboration et présentation* ». Cette norme spécifie que les termes reliés doivent figurer en gras quand ils sont définis et reproduits dans les listes terminologiques normalisées. Cette typographie sert d'indicateur de liens pour générer les nœuds équivalents dans les cartes graphiques conceptuelles correspondantes.

Le travail des experts du WG1 consiste, donc, à reproduire ces liaisons et à en proposer d'autres pour combler les lacunes observées entre les concepts. Cette méthode a l'avantage de donner une vue globale du champ sémantique du domaine et d'accélérer son évolution et son appropriation.

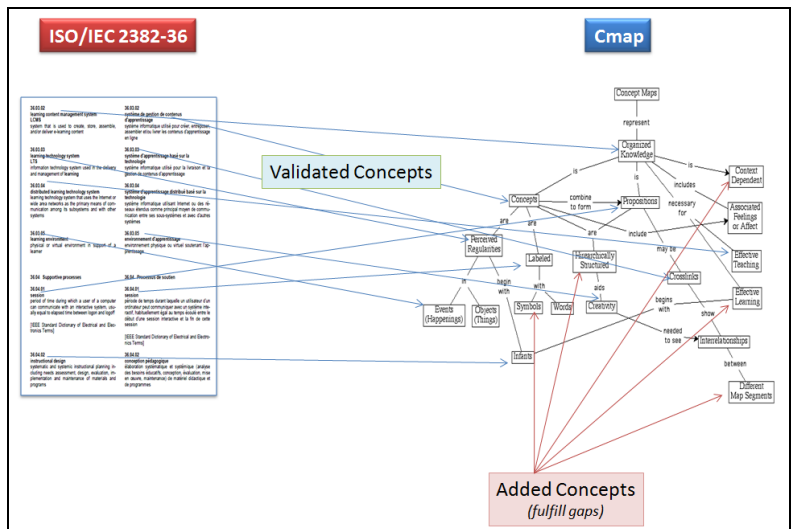

*Fig. 3 : Le passage d'une liste terminologique à une représentation sémantique*

Depuis la session d'Umeå en Suède (septembre 2009), la délégation coréenne a proposé, à travers le NIPA (*National IT Industry Promotion Agency*), d'héberger un serveur *Cmap Tools* et d'aider à la production de cartes graphiques conceptuelles pour le WG1. Une résolution dans ce sens a été retenue pour demander au NIPA de préparer aussi un environnement collaboratif en version Wiki et un guide d'utilisation pour faciliter l'accès des experts du WG1 à cette activité de graphes conceptuels.

Ce travail sera complété par le développement d'une application par le WG1 permettant de produire des cartes conceptuelles conformes aux spécifications appliquées par le WG4 pour produire une carte conceptuelle pour le MLR1 (norme ISO/IEC 19788-1, clause 3, « definitions »).

Pour résumer l'approche globale du mode de fonctionnement du WG1, le travail terminologique dans le domaine de l'e-Learning est conduit sur deux axes : d'une part, il est question de produire des listes de termes et de définitions pour un vocabulaire e-Learning normalisé, et d'autre part il s'agit de produire des graphes conceptuels à partir de ces termes et des définitions retenus (et autres proposés) pour mieux maîtriser la sémantique du domaine. Le tout s'accomplit dans un processus transversal d'harmonisation des procédures et des contenus en se basant sur les référentiels en vigueur et les différentes sources terminologique reconnues dans le domaine.

### *Un cadre d'appui au travail du SC36*

Pour accomplir ces tâches, le WG1 fait appel à des ressources externes qui renforcent ses activités sur le plan terminologique. Il s'agit de ressources en accès libre, de projets et de structures associées dans le cadre de partenariats internationaux.

Au sein de l'ISO, le SC36 coopère étroitement avec l'ITVMT (*Information Technology Validation and Maintenance Team*), créé en 2008 comme organe du JTC1 auprès duquel il gère le suivi et le contrôle des activités relatives aux ressources terminologiques de l'ISO. L'ITVMT gère particulièrement les différentes parties de la norme « *ISO/IEC 2382-x* » conçue pour définir les vocabulaires de plusieurs domaines reliés aux TIC. Ce groupe collabore étroitement avec le groupe de la base de données conceptuelle (ISO/CDB) et le bureau de la traduction du gouvernement du Canada qui gère la banque de données *TERMIUM Plus®*.

L'ISO/CDB est une base de données conceptuelle mise en œuvre pour la première fois en 2009. Elle offre une nouvelle approche de gestion de la qualité dans le développement des ressources de contenus structurés (du moins au niveau de la sémantique lexicale) et dans l'intégration et l'interopérabilité de ces contenus. Elle a été développée pour mieux gérer une terminologie normalisée et des ressources de contenu interopérables. Elle permet une recherche de concepts dans trois importantes catégories de données : concepts (termes et définitions), symboles graphiques et codes (de pays, monnaies, langues et codets).

La recherche des termes et des définitions est un bon exemple de la valeur ajoutée de la base conceptuelle de l'ISO. Les experts peuvent y trouver toutes les occurrences de termes et de définitions provenant de toutes les normes et tous les projets de l'ISO. L'accès aux connaissances et leur partage sont ainsi facilités ; ce qui contribue à améliorer la qualité du contenu et à éviter les doubles emplois ou les incohérences.

*TERMIUM Plus®* est une banque de données terminologiques et linguistiques reconnue à l'échelle internationale, créée par le bureau de la traduction du gouvernement du Canada (Bureau de la traduction (2006). Elle sert de ressources incontournables pour la traduction d'expressions et des mots dans beaucoup de secteurs d'activité. Elle contient près de 4 millions de termes en anglais, en français et en espagnol.

*TERMIUM Plus®* constitue un outil de validation des choix terminologiques réalisés et fournit les éléments nécessaires pour harmoniser les termes et les définitions retenus pour les nouveaux textes de normes élaborés par le SC36.

Le SC36 compte aussi sur ses liaisons internes. L'AUF et l'Alliance Cartago, deux liaisons auprès du SC36 et du WG1, proposent depuis quelque temps de développer un environnement collaboratif de production terminologique et sémantique du domaine. Ces deux partenaires ont opté pour un environnement de développement autour de l'application « *Greenstone* », une solution *Open source* néozélandaise, pour la construction et la distribution de collections de bibliothèques numériques (Lim, 2002). Le projet consiste à produire un module terminologique complémentaire qui sera hébergé sur les serveurs de l'AUF.

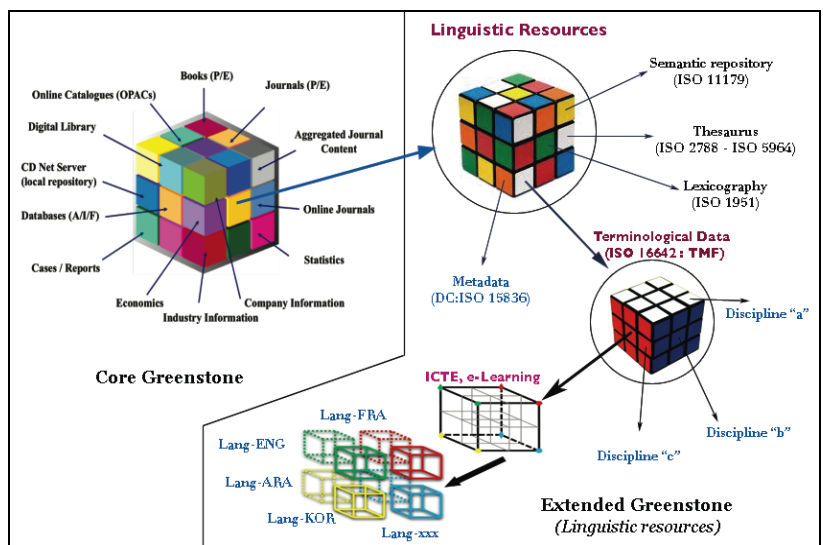

*Fig. 4 : L'extension terminologique du noyau Greestone*

## 6. Perspectives

Pour conclure, nous pensons que les enjeux d'une normalisation du domaine des terminologies e-Learning sont cruciaux pour les communautés académiques et de recherche. Nous avons conscience bien évidemment que de nombreuses questions restent en suspend notamment pour ce qui est des réseaux sémantiques universitaires. Ceux-ci restent à construire en vrai grandeur et nous paraissent comme indispensables à l'âge de la mondialisation. Comment rallier concrètement plusieurs acteurs du monde universitaire dans un projet fédérateur pour un réseau sémantique éducatif ? Par quels moyens serait-il possible de créer un consensus général sur l'intérêt de la normalisation pour le fonctionnement pédagogique, administratif et organisationnel des institutions universitaires ? Qu'en sera-t-il, par exemple, de la contribution des structures universitaires dans la construction des réseaux sémantiques du Web 3.0 ? Comment assurera-t-on un lissage éditorial uniforme des ressources pédagogiques ? Quelles solutions faudra-t-il concevoir pour faire coexister et coopérer des cultures linguistiques ou des écritures très disparates dans la construction d'un réseau sémantique universel et transparent ? Il nous parait évident que l'étude de ces questions est difficile à conduire, mais cela constitue pour nous une raison supplémentaire pour ouvrir des pistes complémentaires de réflexions et de recherche. La production des normes internationales continuera à croître, mais s'il n'y a pas derrière une appropriation par les usagers, le fait de continuer à faire des normes deviendra rapidement contre productif. Un champ d'investigation important s'ouvre ainsi à nous pour tenter d'approfondir notre compréhension des usages et des usagers dans le monde de la terminologie et de la sémantique normalisée. Ceci fera certainement l'objet de l'une de nos perspectives de recherche dans ce domaine nouveau.

## 7. Bibliographie

Polguère A., *Lexicologie et sémantique lexicale : notions fondamentales.* PUM, 2003, 260 p., ISBN : 2760618609